\documentclass[12pt]{article}

\usepackage{amsbsy,amsmath,amsthm,amssymb}
\usepackage{graphics}
\usepackage[demo]{graphicx}
\usepackage{setspace}
\usepackage{natbib}
\usepackage{fullpage}
\usepackage{url}
\usepackage{multirow}

\theoremstyle{plain}

\theoremstyle{definition}

\theoremstyle{remark}

\newcommand{\Real}{\mathbb{R}}

\DeclareMathOperator{\Cs}{\mathcal{C}}
\DeclareMathOperator{\Ss}{\mathcal{S}}
\DeclareMathOperator{\Ls}{\mathcal{L}}

\DeclareMathOperator{\Vs}{\mathcal{V}}

\DeclareMathOperator{\card}{card}

\newcommand{\argmin}{\qopname\relax m{arg\,min}}

\makeatletter
\renewcommand{\section}{\@startsection {section}{1}{\z@}%
    {-3.5ex \@plus -1ex \@minus -.2ex}%
    {2.3ex \@plus.2ex}%
    {\centering\normalfont\Large\bfseries\uppercase}}
\makeatother

\begin{document}

\begin{center}
\LARGE \bf Automatic Response Category Combination in Multinomial Logistic Regression
\end{center}
\begin{center}
Bradley S. \textsc{Price}, Charles J. \textsc{Geyer},
and Adam J. \textsc{Rothman}
\end{center}

\newcommand{\foo}{\thefootnote}
\renewcommand{\thefootnote}{\relax}
\footnotetext{Bradley S. Price, Management Information Systems Department, West Virginia University
    (E-Mail: \emph{brad.price@mail.wvu.edu)}.
    Charles J. Geyer, School of Statistics, University of Minnesota
    (E-mail: \emph{geyer@umn.edu}).
    Adam J. Rothman, School of Statistics, University of Minnesota
    (E-mail: \emph{arothman@umn.edu}).}
\renewcommand{\thefootnote}{\fred}

\begin{abstract}
 We propose a penalized likelihood method that simultaneously
fits the multinomial logistic regression
model and combines subsets of the response categories.
The penalty is nondifferentiable when pairs of columns in
the optimization variable are equal. This encourages
pairwise equality of these columns in the estimator, which corresponds to response category
combination. We use an alternating direction method of multipliers algorithm
to compute the estimator and we discuss the algorithm's convergence.
Prediction and model selection are also addressed.

\par
\medskip
\noindent \textbf{Key Words:} Multinomial logistic regression; 
Fusion penalty; Response category reduction.
\end{abstract}

\section{Introduction}

\begin{sloppypar}
We propose a new way to fit the multinomial logistic regression model.
Let $x_{i}=(1,x_{i2},\ldots, x_{ip})' \in\mathbb{R}^p$  
be the non-random values of the  predictors for the $i$th subject 
and let $y_{i} = (y_{i1},\ldots,y_{iC})' \in \mathbb{R}^C$ be the 
observed response category counts for the $i$th subject $(i=1,
\ldots,n)$.  The model assumes that $y_i$ is a realization of the random vector
 \begin{equation}
 \label{MultiGen}
 Y_i \sim {\rm Multinomial}\left(n_i, \pi^*_{i1},\ldots, \pi^*_{iC}\right), \quad i=1,\ldots, n
 \end{equation}
where $n_i$ is the index of the multinomial experiment for the $i$th subject
and $\pi^*_{ij}$ is the unknown probability 
that the response is category $j$ for the $i$th subject 
$(i,j) \in \{1, \ldots, n\} \times \Cs$, where $\Cs=\{1,2,\ldots,C\}$.
The model also assumes that $Y_1,\ldots, Y_n$ are independent.
Using the baseline category parameterization, 
\begin{equation}
\label{logit}
\pi^*_{ij}=\frac{\exp(x_{i}'\beta^*_j)}{\sum_{m=1}^C\exp(x_{i}'\beta^*_m)}, \, \, \qquad j \in \Cs,
\end{equation}
where $\beta_{1}^{*}, \ldots, \beta_{C-1}^{*}$ are unknown regression 
coefficient vectors and $\beta_{C}^{*} = \vec{0}$.  
Other constraints could be used to
make $\beta^*_1, \ldots, \beta_{C}^*$ identifiable, e.g. 
\citet{zhu04} used $\sum_{j=1}^C \beta^*_j=0$.
\end{sloppypar}

Let $\ell: \mathbb{R}^{p} \times \cdots \times \mathbb{R}^{p} 
\rightarrow \mathbb{R}$ be the log likelihood with additive terms
that do not contain parameters dropped:
\begin{equation}
\label{likeUnPen}
\ell(\beta_1, \ldots, \beta_{C-1})=
\sum_{i=1}^n\sum_{j=1}^{C-1} y_{ij}x_{i}'\beta_{j}-\sum_{i=1}^n n_i\log\left\{1+\sum_{r=1}^{C-1}
\exp\left( x_{i}' \beta_r\right)\right\}.
\end{equation}
There are a total of $p(C-1)$ unknown parameters.  
For more information on the model see \citet[Chapter~8]{agresti02}.

Shrinkage or regularized estimation is natural in this setting when $p(C-1)$ is large.  
\citet{zhu04} proposed ridge-penalized likelihood estimation,
and \citet{vincent14} proposed sparse group-lasso penalized likelihood estimation that encourages
estimates of the matrix $(\beta_{1},\ldots, \beta_{C-1})\in\mathbb{R}^{p\times C-1}$
that have 0's in all entries of some rows.  
These rows correspond to explanatory variables
that are estimated to be irrelevant. A similar procedure was studied by \citet{simon14}.

Interpreting the regression coefficients when the response has three or more 
categories is difficult because the coefficient 
values depend on the baseline category choice (or depend on another 
constraint employed to make the parameters identifiable).
Interpretation in the special case of binomial logistic regression ($C=2$)
is much simpler.  In some applications, we can create a two-category response
by subjectively grouping the original response categories, e.g. if the $j$th
response category was of interest, one could create two groups of categories $\{j\}$
and $\Cs\setminus \{j\}$, and perform binomial logistic regression.  Without
subjective information, we can view response category grouping as a model selection 
problem, where the selected submodel with grouped categories 
fits the counts in our data nearly as well as the full model.

We propose a penalized likelihood procedure that uses fusion 
penalties to encourage fitted models 
with grouped response categories.  Model selection is addressed 
using both $K$-fold cross validation and AIC.  Through simulation, we show that in settings where the true response categories are grouped, our method 
using cross validation for tuning 
parameter selection performs better than competitors at  
predicting the true response category probabilities. We also show that
in certain settings, our method using AIC for model selection 
excels at selecting the true grouped response categories.

\section{Method}
\label{method}

We propose the penalized likelihood estimates defined by
\begin{align} \label{ObjFun}
(\hat\beta_1,\ldots, \hat\beta_{C})  = 
\argmin_{(\beta_1,\ldots, \beta_{C})\in\mathbb{R}^p\times\cdots\times\mathbb{R}^p}
& \left\{  -\ell(\beta_1,\ldots, \beta_{C-1}) + \lambda \sum_{(j,m)\in\mathcal{L}} |\beta_j - \beta_m|_2 \right\}\\
& {\rm subject} \ {\rm to} \ \beta_C=\vec{0}, \nonumber
\end{align}
where $\mathcal{L}$ is a user-selected subset of $\Ss=\{(a,b)\in \Cs \times\Cs : a > b\}$,
$\lambda \geq 0$ is a tuning parameter, and
$|\cdot |_2$ is the vector 2-norm.
Since the penalty only depends on differences of $\beta$ vectors,
the estimated response category probabilities
are invariant to the choice of the baseline category for a given $\Ls$.

To exploit a natural ordering (if any) of 
the response categories, one could set  $\Ls=\{(1,2), (2,3), \ldots, (C-1, C)\} $. Without prior information about the 
similarity between response categories, one could set $\Ls=\Ss$.

Suppose that $\Ls=\Ss$. The objective function in \eqref{ObjFun} is non-differentiable 
$(\beta_1,\ldots, \beta_{C})$ has 
at least one pair of equal components, $\beta_m = \beta_j$, $m\neq j$.  This
encourages an increasing number of 
pairs of equal element vectors in $(\hat\beta_1,\ldots, \hat\beta_{C})$  
as $\lambda$ increases.  If $\hat\beta_m = \hat\beta_j$, then the 
corresponding estimated
response category probabilities are also equal.
In effect, response categories $j$ and $m$ are combined.

If $\lambda$ is 
sufficiently large, then 
$\hat\beta_1=\cdots=\hat\beta_C = 0$.  In this uninteresting edge-case, 
all $\hat\pi_{ij}$'s are equal to $1/C$.  

We select $\lambda$ using $K$-fold cross validation maximizing the validation log-likelihood.  Let $\Vs_k$ be the set of indices in 
the $k$-th fold.
Specifically we maximize the function $Q$, defined by

\begin{equation}
\label{vl}
Q(\lambda)=\sum_{k=1}^K\sum_{i \in \Vs_k}\sum_{l=1}^C y_{il}\log\left(\widehat{\pi}_l^{(-\Vs_k)}(x_i,\lambda)\right),
\end{equation}
where $y_{il}$ is the observed  
count for  response category $l$ for validation observation $i$, and
where $\widehat{\pi}^{(-\Vs_q)}_l(x_i,\lambda)$ is the estimated probability
validation observation $i$, with predictor values $x_i$, has response 
category $l$ using estimates produced from the training set that omits data
with indices in $\Vs_q$ using tuning parameter $\lambda$.
We propose an alternative method of tuning parameter selection using AIC
in Section~\ref{TuneReg}.

Our penalty builds on
the ideas behind the group lasso penalty \citep{bakin99, yuan06} and the fused lasso 
penalty \citep{TibshiraniEtAl05}.  \citet{Alaiz13} developed the grouped fused lasso
penalty, but their work only fused adjacent groups
and was designed for single-response penalized least-squares regression.  Both  \citet{hocking} and \citet{CheChiRan2015} 
investigate similar penalties in the context of clustering.

\section{Algorithm}

 We propose to solve \eqref{ObjFun} using an alternating direction method of multipliers algorithm (ADMM).  
 \citet{Boyd10} provide an introduction and  description of the ADMM algorithm.  The proposed ADMM algorithm solves the 
 optimization in \eqref{ObjFun} by solving the
equivalent constrained optimization:

\begin{align}
\label{Reform}
 \argmin_{\beta_j \in \Real^p,\, j=1,\ldots, C, Z_{j,m} \in\Real^{p},\, (j,m)\in \Ls} &\left\{-\ell(\beta_1,\ldots,\beta_{C-1})+
\lambda\sum_{(j,m)\in \Ls}\rvert Z_{j,m}\lvert_2
\right\},\\
&\mbox{subject to} \,\, \beta_C=\vec{0}\, \mbox{and}\,  Z_{j,m}=\beta_j-\beta_m, (j,m) \in \Ls.\nonumber
\end{align}
The optimization in \eqref{Reform} can be written as a special case of equation 3.1 of \citet{Boyd10}.  Define $\beta=(\beta_1,
\ldots,\beta_{C-1})$, $Z=\{Z_{j,m}\}_{(j,m) \in \Ls}$ and $W=\{W_{j,m}\}_{(j,m) \in \Ls}$, where $W_{j,m}$ is the Lagrange 
multiplier associated with the constraint on $Z_{j,m}$.  The augmented Lagrangian is
\begin{align*}
L_\rho(\beta, Z,W)=&-\ell(\beta)\\
&+\sum_{(j,m) \in \Ls}\left\{ \lambda|Z_{j,m}|_2+W_{j,m}^T(\beta_j-\beta_m-Z_{j,m})+
\frac{\rho}{2}|\beta_j-\beta_m-Z_{j,m}|_2^2\right\},
\end{align*}
where $\rho>0$ is the augmented Lagrangian parameter.  
We now apply equations 3.2, 3.3, and 3.4 of \citet{Boyd10} to obtain the ADMM update equations for solving \eqref{Reform}:
\begin{align}
\label{Ridge}
\beta^{(k+1)}&=\argmin_{\beta_j \in \Real^p\, j=1,\ldots, C-1} L_{\rho}(\beta,Z^{(k)},W^{(k)})\\
\label{ZUpdate}
 Z^{(k+1)}&=\argmin_{Z_{j,m} \in \Real^{p}, (j,m) \in \Ls}  L_{\rho}(\beta^{(k+1)},Z,W^{(k)}), \\
 \label{UUpdated}
 W_{j,m}^{(k+1)}&=W_{j,m}^{(k)}+\rho(\beta^{(k+1)}_j-\beta_m^{(k+1)}-Z^{(k+1)}_{j,m}),\,  (j,m)\in \Ls,
\end{align}
where a superscript of $(k)$ denotes the $k$th iterate.
 
We propose to solve \eqref{Ridge} with blockwise coordinate descent,  where 
$\beta_1,\ldots,\beta_{C-1}$ are the blocks.  In each block update,
we use Newton's method, which converges because these subproblems have
strongly convex objective functions.
Solving \eqref{Ridge} is the most computationally expensive step.

The optimization  in \eqref{ZUpdate}  decouples into $\card(\Ls)$ optimization problems that can be solved
in parallel with the following closed form solutions:
\begin{equation}
\label{ZSol}
Z_{j,m}^{(k+1)}=\left(\beta_j^{(k+1)}-\beta_m^{(k+1)}+\frac{1}{\rho}W_{j,m}^{(k)}\right)\left(1-\frac{\lambda}{\rho\lvert 
\beta_j^{(k+1)}-\beta_m^{(k+1)}+\frac{1}{\rho}W^{(k)}_{j,m} \rvert_2}\right)_+, \,(j,m) \in \Ls,
\end{equation} 
where $(a)_+=\max(a,0)$.  This solution is a special case of solving the group lasso penalized least squares problem 
\citep{yuan06}. 

One could show  convergence of the proposed ADMM algorithm by using the theory developed by \citet{ADMMConv}. This proof 
relies on the assumption that the undirected graph defined by the vertex set $\Cs$ and the edge set $\Ls$, is connected.   In 
practice the convergence tolerance and step size $\rho$ can be determined 
by equation 3.12 and 3.13 respectively in \citet{Boyd10}.

The final iterate $\beta^{(K)}$ will not have pairs of equal component vectors, but subsets of its pairs of component vectors will 
be very similar.  We define our final estimates by
\begin{equation}
\label{FinalEst}
\hat{\beta}_j=\frac{\sum_{m\in \Cs\setminus\{j\}} \beta_m^{(K)}I(Z^{(K)}_{j,m}=\vec{0})+\beta^{(K)}_j}{\sum_{m \in \Cs
\setminus\{j\}}I(Z^{(K)}_{j,m}=\vec{0})+1}, \, j=1,\ldots, C-1,
\end{equation}
and $\hat{\beta}_C=\vec{0}$.  
We set $\hat{\beta}_1,\ldots,\hat{\beta}_{C-1}$  equal to $0$ if the 2-norm of the vector is less than $10^{-8}$. A
similar approach was taken by \cite{JGL}.

\section{Prediction Performance Simulations}
\label{predpre}
\subsection{Competitors}
\label{comp_methdos}

We present simulation studies that compare the prediction performance of our proposed method, which we call
group fused multinomial logistic regression (GFMR), 
elastic net penalized multinomial logistic regression (EN),  and group penalized multinomial logistic regression (GMR)  
\citep{GlmNet, simon14}.   
Both methods used for comparison are implemented in the \texttt{glmnet} package in R.   An R package implementing the GFMR methods 
is currently under development.  
\subsection{Data Generating Models}
\label{data_gen}
 In
Sections \ref{pred_sim1} and \ref{pred_sim2}, we present simulation 
results where the data was generated such that $x_i=(1,\tilde{x}_i)'$, where $\tilde{x}_1,\ldots, \tilde{x}_n$ are drawn 
independently from
$N_9(0,I)$.    We set $n_i=1$ for $ i=1,\ldots,n$ and $C=4$.  We consider two settings for $\beta^*$.  
In setting 1, $\beta^*_1=\beta^*_4=\vec{0}$ and $
\beta_2^*=\beta_3^*=\vec{\delta}$.  In setting 2,  $
\beta^*_1=\beta^*_2=\beta^*_3=\vec{\delta}$, and $\beta^*_4=\vec{0}$. We consider  $(n,\delta)\in \{ 50,100\} \times 
\{0.1, 0.25,0.5, 1\}$. The observed responses $y_i,\dots,y_n$ are a realization of $Y_1,\ldots,Y_n$ defined by 
\eqref{MultiGen}.

Results are based on 100 replications.  We consider two choices for $\Ls$: $\Ls=\Ss$ and $
\Ls=\{(1,2), (2,3), (3,4)\}$. The second choice attempts to exploit a natural ordering of the 
response categories.  In each replication, we measure the prediction performance using the 
Kullback-Leibler (KL) divergence between the estimated response category probabilities and the 
true response category probabilities, based on 1000 test observations generated from the same 
distribution as the training data. The KL Divergence is defined by

\begin{equation}
\label{kl}
KL(\hat{\pi},\pi^*)=\frac{1}{1000}\sum_{i=1}^{1000}\sum_{k=1}^4 \log\left(\frac{\widehat{\pi}_k(x_i,\lambda)}{\pi^*_k(x_i)}\right)\widehat{\pi}_k(x_i,\lambda), 
\end{equation}
where $\pi^*_k(x_i)$, and $\widehat{\pi}_k(x_i,\lambda)$  are the true and 
estimated response category probabilities that testing observation $i$ has response category $k$ using tuning parameter $\lambda$.  For both GFMR methods we report the average number of unique regression coefficient vectors observed in the estimator for the 100 replications.  The number of unique regression coefficient vectors estimated by the GFMR methods is defined as the unique vectors in $(\hat{\beta}_1,\ldots,\hat{\beta}_C)$.

Tuning parameters for GFMR are selected from a subset of 
$\{10^{-10},10^{-9.98},\ldots, 10^{9.98},10^{10}\}$. We also computed the best overall KL divergence we could obtain using the candidate values for the tuning parameter.  We call this
the oracle tuned value.  The tuning parameters for the competing methods were selected using 5-fold cross validation minimizing the validation deviance.  For the both EN and GMR, the first tuning parameter was selected by default methods in the \texttt{glmnet} package in R and the second
tuning parameter was selected from the set $\{0,0.01,\ldots,0.99,1\}$. 

\subsection{Results for Setting 1}
\label{pred_sim1}

In Table \ref{tune_sim1}, we present the average KL divergence and the average number of unique regression coefficient vectors estimated by the GFMR method for the simulation using setting 1 when $\Ls=\Ss$.  On average GFMR using validation likelihood had a lower KL divergence than its competitors.  Since there should be 2 unique regression coefficient vectors in the estimator, the results show GFMR using oracle tuning on average is overselecting the number of unique coefficient 
vectors when $\delta=0.25,0.5$, 
and $1$ for both values of $n$.    A similar result occurred for GFMR using validation likelihood for all values of $\delta$ and $n$.  This may indicate that we need
 a different method to select the tuning parameter for GFMR if our interest is in model selection, when $\Ls=\Ss$, rather than prediction of the true response category probabilities where this method performs better than EN and GMR.

The average KL divergence and the average number of unique regression coefficient vectors estimated by GFMR for setting 1 when $\Ls=\{(1,2), (2,3), (3,4)\}$ are 
reported in Table \ref{ord_set1}.  These results show the same pattern that the results in Table \ref{tune_sim1} showed.  While prediction accuracy decreased in this setting, we again see the GFMR using validation likelihood tuning is competitive with GMR. This decrease in prediction performance is expected because response categories are not naturally ordered in this data generating model. We also saw an increase in 
the average number of unique regression coefficient vectors estimated using GFMR with oracle tuning and GFMR with validation likelihood tuning when compared to the setting $\Ls=\Ss$.  

To further investigate the number of unique regression coefficient vectors estimated by GFMR, we present a comparison between this quantity for the case when $\Ls=\Ss$ and $\Ls=\{(1,2), (2,3), (3,4)\}$ in Table \ref{setting1_gcomp}.  We see that GFMR using validation likelihood performed poorly at 
selecting the true number of unique coefficient vectors estimated for both choices of $\Ls$.

\begin{table}
\caption{Results of simulation using setting 1 when $\Ls=\Ss$ .  Standard errors are presented in parenthesis.  The columns Oracle, VL, EN, and GMR report to the average KL divergence for oracle tuned GFMR, GFMR using validation likelihood, EN, and GMR methods respectively.  The column VL-Oracle reports the average difference
for GFMR using validation likelihood and oracle tuned GFMR.  Similar notation is used for comparisons between VL and EN, and VL and GMR.  The columns labeled Oracle Groups, and VL Groups, contain the average number of unique regression coefficient vectors when using oracle tuned GFMR, and GFMR using validation likelihood.}
\label{tune_sim1}
\begin{center}
\scalebox{0.8}{
\begin{tabular}{cc|c|c|c|c|c|c|c|c|c}

  & $\delta$ & Oracle & VL & VL-Oracle & \begin{tabular}{c}Oracle\\ Groups\end{tabular} & \begin{tabular}{c}VL\\ Groups\end{tabular} & EN & VL-EN& GMR& VL-GMR \\
\hline 
\multirow{8}{*}{$n=50$} & 0.10 &\begin{tabular}{c} 0.011\\  (0.001)\end{tabular} & \begin{tabular}{c}0.023 \\ (0.000) \end{tabular} & \begin{tabular}{c}
 0.011\\ (0.002)\end{tabular} & 1.95 & 2.98 &\begin{tabular}{c} 0.050\\ (0.003)\end{tabular} & \begin{tabular}{c} -0.027\\ (0.003)\end{tabular}&\begin{tabular}{c} 0.049\\ (0.003)\end{tabular} &\begin{tabular}{c} -0.027\\ (0.003)\end{tabular} \\ \cline{2-11}
 & 0.25 & \begin{tabular}{c} 0.057 \\ (0.002) \end{tabular} & \begin{tabular}{c} 0.076\\ (0.002)\end{tabular} & \begin{tabular}{c}
 0.019\\ (0.002) \end{tabular} & 2.98 & 3.91 & \begin{tabular}{c} 0.094\\ (0.003) \end{tabular} & \begin{tabular}{c} -0.017\\ (0.004) \end{tabular} &\begin{tabular}{c} 0.091\\ (0.003)\end{tabular} &\begin{tabular}{c} -0.015\\ (0.003)\end{tabular}   \\ \cline{2-11}
  & 0.50& \begin{tabular}{c} 0.137\\ (0.003) \end{tabular} & \begin{tabular}{c} 0.164\\ (0.005) \end{tabular} & \begin{tabular}{c}
0.027\\ (0.003) \end{tabular}  & 3.51 & 4 & \begin{tabular}{c}0.191\\ (0.005)\end{tabular}& \begin{tabular}{c}-0.027\\ (0.004)\end{tabular}&\begin{tabular}{c} 0.188\\ (0.004)\end{tabular} &\begin{tabular}{c} -0.024\\ (0.004)\end{tabular}  \\ \cline{2-11}
  & 1.00 & \begin{tabular}{c} 0.259\\ (0.005)\end{tabular} & \begin{tabular}{c} 0.292\\ (0.005) \end{tabular} & \begin{tabular}{c}
0.033\\ (0.004)\end{tabular} & 4 & 4 & \begin{tabular}{c} 0.399\\ (0.010)\end{tabular} & \begin{tabular}{c}  -0.107\\ (0.009)\end{tabular}&\begin{tabular}{c} 0.399\\ (0.010)\end{tabular}&\begin{tabular}{c} -0.106\\ (0.008)\end{tabular} \\ 
\hline 
\multirow{8}{*}{$n=100$}  & 0.10 & \begin{tabular}{c} 0.011\\ (0.001)\end{tabular} & \begin{tabular}{c} 0.019\\ (0.001) \end{tabular} & \begin{tabular}{c}
 0.008\\ (0.001) \end{tabular} & 2.27 & 3.36 &\begin{tabular}{c} 0.033\\ (0.002)\end{tabular} &\begin{tabular}{c} -0.013\\ (0.001) \end{tabular} &\begin{tabular}{c} 0.032\\ (0.002)\end{tabular}&\begin{tabular}{c} -0.012\\ (0.001)\end{tabular}\\ \cline{2-11}
  & 0.25  & \begin{tabular}{c} 0.050\\ (0.001)\end{tabular} & \begin{tabular}{c} 0.061\\ (0.001) \end{tabular} & \begin{tabular}{c}
 0.008\\ (0.001) \end{tabular} & 2.27 & 3.36 & \begin{tabular}{c} 0.033\\ (0.002)\end{tabular} & \begin{tabular}{c} -0.009\\ (0.001)\end{tabular}&\begin{tabular}{c} 0.069\\ (0.002)\end{tabular}&\begin{tabular}{c} -0.012\\ (0.001)\end{tabular} \\ \cline{2-11}
  & 0.50& \begin{tabular}{c} 0.098\\ (0.002) \end{tabular} & \begin{tabular}{c} 0.114\\ (0.004) \end{tabular} & \begin{tabular}{c}
0.016\\ (0.003) \end{tabular} & 3.86 & 4 & \begin{tabular}{c} 0.141\\ (0.005)\end{tabular} & \begin{tabular}{c}-0.027\\ (0.003)\end{tabular} & \begin{tabular}{c} 0.142\\ (0.004)\end{tabular}& \begin{tabular}{c} -0.027\\ (0.003)\end{tabular} \\ \cline{2-11}
  & 1.00 & \begin{tabular}{c} 0.150\\ (0.004) \end{tabular} & \begin{tabular}{c} 0.164\\ (0.004) \end{tabular} & \begin{tabular}{c}
0.015\\ (0.002)\end{tabular}   & 4 & 4 & \begin{tabular}{c} 0.222\\ (0.006)\end{tabular} & \begin{tabular}{c}-0.058\\ (0.005)\end{tabular} 
 &\begin{tabular}{c} 0.225\\ (0.006)\end{tabular} &\begin{tabular}{c} -0.061\\ (0.005)\end{tabular}
\end{tabular}
}
\end{center}
\end{table}

\begin{table}
\caption{Results of simulation using setting 1 when $\Ls=\{(1,2), (2,3), (3,4)\}$.  The columns Oracle, and VL report to the average KL divergence for oracle tuned GFMR and GFMR using validation likelihood methods respectively.  The column VL-Oracle reports the average difference
for GFMR using validation likelihood and oracle tuned GFMR.  Similar notation is used for comparison VL and GMR.  The columns labeled Oracle Groups, and VL Groups, contain the average number of unique regression coefficient vectors when using oracle tuned GFMR, and GFMR using validation likelihood.}
\label{ord_set1}
\begin{center}
\scalebox{0.8}{
\begin{tabular}{cc|c|c|c|c|c|c}
  & $\delta$ & Oracle & VL & VL-Oracle & Oracle Groups & VL Groups & VL-GMR \\
\hline 
\multirow{8}{*}{$n=50$} & \multirow{2}{*}{0.10} &\begin{tabular}{c}0.0121\\ (0.000)\end{tabular} & \begin{tabular}{c} .025\\ (0.003)\end{tabular} & \begin{tabular}{c} 0.013\\ (0.002)\end{tabular} & 3.1 & 3.1 & \begin{tabular}{c}
-0.024\\ (0.003)
\end{tabular} \\ \cline{2-8}
  &\multirow{2}{*}{0.25} & \begin{tabular}{c}0.065\\ (0.001)\end{tabular} &\begin{tabular}{c} 0.0.086\\ (0.003)\end{tabular} &\begin{tabular}{c} 0.028\\ (0.003)\end{tabular} & 3.4 & 3.25&\begin{tabular}{c} -0.006\\ (0.003)\end{tabular} \\ \cline{2-8}
  & \multirow{2}{*}{0.50} & \begin{tabular}{c}0.156\\ (0.004)\end{tabular} & \begin{tabular}{c} 0.198\\ (0.005)\end{tabular} & \begin{tabular}{c} 0.040\\ (0.005)\end{tabular} & 3.95 & 3.62& \begin{tabular}{c} 0.009\\ (0.004)\end{tabular} \\ \cline{2-8}
  & \multirow{2}{*}{1.00} & \begin{tabular}{c}0.189\\ (0.003)\end{tabular} &\begin{tabular}{c} 0.245\\ (0.003)\end{tabular} & \begin{tabular}{c}0.054\\ (0.001)\end{tabular} & 4 & 3.77& \begin{tabular}{c} -0.017\\ (0.010)\end{tabular} \\
\hline 
\multirow{8}{*}{$n=100$} &\multirow{2}{*}{0.10} & \begin{tabular}{c}0.012\\ (0.000)\end{tabular} & \begin{tabular}{c}0.020\\ (0.002)\end{tabular} &\begin{tabular}{c} 0.008\\ (0.001)\end{tabular} & 2.91 & 3.15 &\begin{tabular}{c} -0.011\\ (0.001)\end{tabular} \\  \cline{2-8}
  & \multirow{2}{*}{0.25} & \begin{tabular}{c}0.058\\ (0.001)\end{tabular} & \begin{tabular}{c}0.069\\ (0.001)\end{tabular} & \begin{tabular}{c} 0.011\\ (0.002)\end{tabular} & 3.65 & 3.41&\begin{tabular}{c} 0.000\\ (0.002)\end{tabular} \\ \cline{2-8}
  & \multirow{2}{*}{0.50} & \begin{tabular}{c}0.107\\ (0.003)\end{tabular} & \begin{tabular}{c} 0.127\\ (0.005)\end{tabular} &\begin{tabular}{c} 0.020\\ (0.004)\end{tabular} & 3.99 & 3.86& \begin{tabular}{c} -0.014\\ (0.003)\end{tabular} \\ \cline{2-8}
  & \multirow{2}{*}{1.00} & \begin{tabular}{c}0.156\\ (0.004)\end{tabular} & \begin{tabular}{c}0.191\\ (0.005)\end{tabular} & \begin{tabular}{c}0.035\\ (0.004)\end{tabular} & 4 & 3.99&\begin{tabular}{c} -0.034\\ (0.003)\end{tabular}
\end{tabular} 
}
\end{center}
\end{table}

\begin{table}
\caption{Number of replications out of 100 that produced the number of unique regression coefficient vectors in setting 1. Ordered penalty set references that the penalty set used is $\Ss$, while unordered penalty set references $\Ls=\{(1,2), (2,3), (3,4)\}$. }
\label{setting1_gcomp}
\begin{center}
\scalebox{.8}{
\begin{tabular}{cc|c|c|c|c|c}
  & $\delta$ & Penalty Set  & 1 Vector & 2 Vectors & 3 Vectors & 4 Vectors \\
\hline 
\multirow{8}{*}{$n=50$} & \multirow{2}{*}{0.10} & Ordered & 0 & 6 & 78 & 16 \\ 
  &   & Unordered & 64 & 6 & 1 & 29 \\ \cline{2-7}
  & \multirow{2}{*}{0.25} & Ordered & 0 & 5 & 65 & 30 \\
  &   & Unordered & 40 & 3 & 3 & 54 \\  \cline{2-7}
  & \multirow{2}{*} {0.50} & Ordered & 0 & 1 & 36 & 63 \\
  &   & Unordered & 16 & 0 & 1 & 83 \\  \cline{2-7}
  &  \multirow{2}{*}{1.00} & Ordered & 0 & 0 & 12 & 88 \\
  &   & Unordered & 0 & 0 & 0 & 100 \\ 
\hline  
\multirow{8}{*}{$n=10$} & \multirow{2}{*}{0.10} & Ordered & 0 & 8 & 68 & 24 \\
  &   & Unordered & 49 & 12 & 2 & 37 \\ \cline{2-7}
  &  \multirow{2}{*}{0.25} & Ordered & 0 & 1 & 57 & 42 \\
  &   & Unordered & 24 & 3 & 4 & 69 \\ \cline{2-7}
  &  \multirow{2}{*}{0.50} & Ordered & 0 & 0 & 14 & 86 \\
  &   & Unordered & 4 & 1 & 0 & 95 \\ \cline{2-7}
  & \multirow{2}{*}{1.00} & Ordered & 0 & 0 & 1 & 99 \\
  &   & Unordered & 0 & 0 & 0 & 100
\end{tabular} 
}
\end{center}
\end{table}

\subsection{Results of Setting 2}
\label{pred_sim2}

In Table \ref{tune_sim2}, we present the average KL divergence and the average 
number of unique regression coefficient vectors estimated by the GFMR method for the simulation using setting 2 when 
$\Ls=\Ss$.  Similar to the patterns shown in setting 1 when $\Ls=\Ss$,  on average the GFMR using validation likelihood tuning 
has a lower KL divergence than both EN and the GMR.  For all values of $n$ and $\delta$, with the exception of $(\delta, n)=(0.1,
50)$, both the oracle tuned GFMR and GFMR using validation likelihood tuning overselects the number 
of unique coefficient vectors.  Again this may indicate the need for a different method for tuning parameter selection if our 
interest is model selection.  

In Table \ref{ord_set2}, we present the average KL divergence and the average number of unique regression coefficients 
estimated by GFMR for the simulation using setting 2 when $\Ls=\{(1,2), (2,3), (3,4)\}$.
We see a similar result to the pattern observed Table \ref{ord_set2}, but the prediction accuracy here was better.  This is 
expected because the response categories have a natural ordering in this data generating model.    

To further investigate this comparison of unique regression coefficient vectors, Table \ref{setting2_gcomp} presents a 
comparison between the number of unique regression coefficient vectors estimated by the GFMR method in each of the 100 
replications for the case when $\Ls=\Ss$ and the case when $\Ls=\{(1,2), (2,3), (3,4)\}$.  Just as in Section \ref{pred_sim1}, 
these results show that GFMR using validation likelihood performs poorly at selecting the true number of unique coefficient 
vectors for both penalty sets, but again on average predicts the true response category probabilities better than EN and GMR.

\begin{table}
\caption{Results of simulation using setting 2 when $\Ls=\Ss$ .  Standard errors are presented in parenthesis.  The columns 
Oracle, VL, EN, and GMR report to the average KL divergence for oracle tuned GFMR, GFMR using validation likelihood, EN, and GMR 
methods respectively.  The column VL-Oracle reports the average difference
for GFMR using validation likelihood and oracle tuned GFMR.  Similar notation is used for comparisons between VL and EN, and VL 
and GMR.  The columns labeled Oracle Groups, and VL Groups, contain the average number of unique regression coefficient 
vectors when using oracle tuned GFMR, and GFMR using validation likelihood.}
\label{tune_sim2}
\begin{center}
\scalebox{0.8}{
\begin{tabular}{cc|c|c|c|c|c|c|c|c|c}

  & $\delta$ & Oracle & VL & VL-Oracle & \begin{tabular}{c}Oracle\\ Groups\end{tabular} & \begin{tabular}{c}VL\\ Groups\end{tabular} & EN & VL-EN& GMR &VL-GMR \\
\hline 
\multirow{8}{*}{$n=50$} & 0.10 &\begin{tabular}{c} 0.008\\  (0.000)\end{tabular} & \begin{tabular}{c}0.023 \\ (0.003) \end{tabular} & \begin{tabular}{c}
 0.014\\ (0.003)\end{tabular} & 1.99 & 2.81 &\begin{tabular}{c} 0.044\\ (0.003)\end{tabular} & \begin{tabular}{c} -0.022\\ (0.002)\end{tabular} &\begin{tabular}{c} 0.043\\ (0.003)\end{tabular}&\begin{tabular}{c} -0.023\\ (0.003)\end{tabular} \\  \cline{2-11}
 & 0.25 & \begin{tabular}{c} 0.043 \\ (0.001) \end{tabular} & \begin{tabular}{c} 0.056\\ (0.002)\end{tabular} & \begin{tabular}{c}
 0.013\\ (0.002) \end{tabular} & 2.45 & 3.75 & \begin{tabular}{c} 0.083\\ (0.003) \end{tabular} & \begin{tabular}{c} -0.026\\ (0.003) \end{tabular} &\begin{tabular}{c} 0.081\\ (0.003)\end{tabular} &\begin{tabular}{c} -0.025\\ (0.003)\end{tabular}  \\ \cline{2-11}
  & 0.50& \begin{tabular}{c} 0.118\\ (0.003) \end{tabular} & \begin{tabular}{c} 0.146\\ (0.003) \end{tabular} & \begin{tabular}{c}
0.026\\ (0.002) \end{tabular}  & 3.07 & 4 & \begin{tabular}{c}0.169\\ (0.003)\end{tabular}& \begin{tabular}{c}-0.023\\ (0.004)\end{tabular} &\begin{tabular}{c} 0.167\\ (0.003)\end{tabular}&\begin{tabular}{c} -0.021\\ (0.003)\end{tabular}  \\ \cline{2-11}
  & 1.00 & \begin{tabular}{c} 0.225\\ (0.006)\end{tabular} & \begin{tabular}{c} 0.243\\ (0.007) \end{tabular} & \begin{tabular}{c}
0.017\\ (0.004)\end{tabular} & 4 & 4 & \begin{tabular}{c} 0.364\\ (0.008)\end{tabular} & \begin{tabular}{c}  -0.120\\ (0.008)\end{tabular}&\begin{tabular}{c} 0.372\\ (0.009)\end{tabular}&\begin{tabular}{c} -0.120\\ (0.008)\end{tabular} \\ 
\hline 
\multirow{8}{*}{$n=100$} & 0.10 & \begin{tabular}{c} 0.008\\ (0.000)\end{tabular} & \begin{tabular}{c} 0.014\\ (0.001) \end{tabular} & \begin{tabular}{c}
 0.006\\ (0.001) \end{tabular} & 2.04 & 3.39 &\begin{tabular}{c} 0.027\\ (0.001)\end{tabular} &\begin{tabular}{c} -0.012\\ (0.001) \end{tabular}&\begin{tabular}{c} 0.027\\ (0.002)\end{tabular}&\begin{tabular}{c} -0.012\\ (0.001)\end{tabular} \\ \cline{2-11}
  & 0.25  & \begin{tabular}{c} 0.037\\ (0.001)\end{tabular} & \begin{tabular}{c} 0.048\\ (0.001) \end{tabular} & \begin{tabular}{c}
 0.010\\ (0.001) \end{tabular} & 2.90 & 3.91 & \begin{tabular}{c} 0.060\\ (0.002)\end{tabular} & \begin{tabular}{c} -0.012\\ (0.002)\end{tabular} & \begin{tabular}{c} 0.059\\ (0.002)\end{tabular} &\begin{tabular}{c} -0.012\\ (0.002)\end{tabular} \\ \cline{2-11}
  & 0.50& \begin{tabular}{c} 0.076\\ (0.002) \end{tabular} & \begin{tabular}{c} 0.089\\ (0.003) \end{tabular} & \begin{tabular}{c}
0.012\\ (0.002) \end{tabular} & 3.85 & 4 & \begin{tabular}{c} 0.116\\ (0.002)\end{tabular} & \begin{tabular}{c}-0.027\\ (0.002)\end{tabular} & \begin{tabular}{c} 0.115\\ (0.003)\end{tabular} &\begin{tabular}{c} -0.026\\ (0.002)\end{tabular} \\ \cline{2-11}
  & 1.00 & \begin{tabular}{c} 0.124\\ (0.004) \end{tabular} & \begin{tabular}{c} 0.131\\ (0.004) \end{tabular} & \begin{tabular}{c}
0.006\\ (0.001)\end{tabular}   & 4 & 4 & \begin{tabular}{c} 0.201\\ (0.007)\end{tabular} & \begin{tabular}{c}-0.070\\ (0.005)\end{tabular} &\begin{tabular}{c} 0.204\\ (0.006)\end{tabular}&\begin{tabular}{c} -0.073\\ (0.005)\end{tabular}
\end{tabular} 
}
\end{center}
\end{table}

\begin{table}  
\caption{Results of simulation using setting 2 when $\Ls=\{(1,2), (2,3), (3,4)\}$ .  The columns Oracle, and VL report to the 
average KL divergence for oracle tuned GFMR and GFMR using validation likelihood methods respectively.  The column VL-Oracle 
reports the average difference
for GFMR using validation likelihood and oracle tuned GFMR.  Similar notation is used for comparison VL and GMR.  The columns 
labeled Oracle Groups, and VL Groups, contain the average number of unique regression coefficient vectors when using oracle 
tuned GFMR, and GFMR using validation likelihood.}

\label{ord_set2}
\begin{center}
\scalebox{0.8}{
\begin{tabular}{cc|c|c|c|c|c|c}
  & $\delta$ & Oracle & VL & VL-Oracle & Oracle Groups & VL Groups &VL-GMR \\
\hline 
\multirow{8}{*}{$n=50$} & 0.10 & \begin{tabular}{c}0.008\\ (0.003)\end{tabular} & \begin{tabular}{c}0.023\\ (0.004)\end{tabular} &\begin{tabular}{c} 0.015\\ (0.003)\end{tabular} & 2.87 & 2.11& \begin{tabular}{c} -0.025\\ (0.003)\end{tabular}\\ \cline{2-8}
  & 0.25 & \begin{tabular}{c} 0.040\\ (0.000)\end{tabular} & \begin{tabular}{c} 0.057\\ (0.002)\end{tabular} & \begin{tabular}{c}0.016\\ (0.001)\end{tabular} & 3.17 & 3.56&\begin{tabular}{c} -0.024\\ (0.003)\end{tabular} \\ \cline{2-8}
  & 0.50 & \begin{tabular}{c}0.106\\ (0.003)\end{tabular} & \begin{tabular}{c} 0.135\\ (0.004)\end{tabular} & \begin{tabular}{c} 0.029\\ (0.003) \end{tabular} & 3.79 & 3.56 & \begin{tabular}{c} -0.032\\ (0.004)\end{tabular} \\ \cline{2-8}
  & 1.00 & \begin{tabular}{c} 0.189\\ (0.003)\end{tabular} & \begin{tabular}{c} 0.245\\ (0.003)\end{tabular} & \begin{tabular}{c} 0.054\\ (0.001)\end{tabular} & 4 & 3.77 & \begin{tabular}{c} -0.128\\ (0.010)\end{tabular} \\
\hline
\multirow{8}{*}{$n=100$} & 0.10 & \begin{tabular}{c} 0.008\\ (0.000)\end{tabular} & \begin{tabular}{c}0.015\\ (0.001)\end{tabular} &\begin{tabular}{c} 0.007\\ (0.001)\end{tabular} & 2.91 & 3.15 & \begin{tabular}{c} -0.011\\ (0.001)\end{tabular} \\ \cline{2-8}
  & 0.25 & \begin{tabular}{c} 0.035\\ (0.001)\end{tabular} &\begin{tabular}{c} 0.046\\ (0.002)\end{tabular} & \begin{tabular}{c} 0.011\\ (0.002)\end{tabular} & 3.43 & 3.35 & \begin{tabular}{c} -0.013\\ (0.002)\end{tabular} \\ \cline{2-8}
  & 0.50 & \begin{tabular}{c} 0.064\\ (0.003)\end{tabular} & \begin{tabular}{c} 0.073\\ (0.003)\end{tabular} & \begin{tabular}{c} 0.009\\ (0.001)\end{tabular} & 3.89 & 3.75 & \begin{tabular}{c} -0.042\\ (0.002)\end{tabular} \\ \cline{2-8}
  & 1 & \begin{tabular}{c} 0.101\\ (0.003)\end{tabular} & \begin{tabular}{c} 0.119\\ (0.004)\end{tabular} & \begin{tabular}{c}0.018\\ (0.002)\end{tabular} & 4 & 3.89& \begin{tabular}{c} -0.085\\ (0.006)\end{tabular}
\end{tabular} 
}
\end{center}
\end{table}

\begin{table}
\caption{Number of replications out of 100 that produced the number of unique regression coefficient vectors in setting 1. 
Ordered penalty set references that the penalty set used is $\Ss$, while unordered penalty set references $\Ls=\{(1,2), (2,3), (3,4)\}$. }
\label{setting2_gcomp}
\begin{center}
\scalebox{0.8}{
\begin{tabular}{cc|c|c|c|c|c}
  & $\delta$ & Penalty Set  & 1 Vector & 2 Vectors & 3 Vectors & 4 Vectors \\
\hline 
\multirow{8}{*}{$n=50$} & \multirow{2}{*}{0.10} & Ordered & 0 & 2 & 83 & 14 \\ 
  &   & Unordered & 63 & 4 & 4 & 29 \\ \cline{2-7} 
  & \multirow{2}{*}{0.25} & Ordered & 0 & 4 & 69 & 27 \\
  &   & Unordered & 49 & 2 & 4 & 45 \\ \cline{2-7}
  & \multirow{2}{*}{0.50} & Ordered & 0 & 2 & 40 & 58 \\
  &   & Unordered & 29 & 3 & 0 & 68 \\ \cline{2-7}
  & \multirow{2}{*}{1.00} & Ordered & 0 & 0 & 22 & 78 \\
  &   & Unordered & 0 & 0 & 0 & 100 \\
\hline 
\multirow{8}{*}{$n=100$} & 0.10 & Ordered & 0 & 7 & 71 & 22 \\ 
  &   & Unordered & 55 & 14 & 3 & 28 \\ \cline{2-7}
  & \multirow{2}{*}{0.25} & Ordered & 0 & 3 & 59 & 38 \\
  &   & Unordered & 33 & 4 & 3 & 60 \\ \cline{2-7}
  & \multirow{2}{*}{0.50} & Ordered & 0 & 0 & 25 & 75 \\
  &   & Unordered & 4 & 1 & 1 & 94 \\ \cline{2-7}
  & \multirow{2}{*}{1.00} & Ordered & 0 & 0 & 11 & 89 \\
  &   & Unordered & 0 & 0 & 0 & 100
\end{tabular} 
}
\end{center}
\end{table}

\section{Two-Step Method for Reducing Response Categories}
\label{TuneReg}

The simulations presented in Section \ref{predpre} show that our method using validation likelihood to select the tuning parameter
performs well at predicting the true response category probabilities  
when compared to EN and GMR.  The same simulations show that 
our method using validation likelihood does not perform well at model 
selection.   To improve 
model selection performance,  we propose an alternative
approach for low dimensional settings.  This is a two-step method 
that solves \eqref{ObjFun}
for an increasing sequence of $\lambda$'s until there are two 
unique vectors in $(\hat{\beta}_1,\ldots,\hat{\beta_C})$ .  
This sequence provides a set of candidate models with different 
response categories from which we will select the best by 
refitting using unpenalized maximum likelihood and computing the 
AIC.   The selected model is the 
candidate model with the minimum AIC.  We also compute the AIC for the edge case where all probabilities are equal to $1/C$.

We present a simulation to show the merits of this two-step 
approach to select the 
correct combined category model in multinomial logistic regression. 
The explanatory variables values are generated using the same 
procedure described in Section \ref{data_gen}.  In this simulation,
$\beta^*_1=-\vec{\delta}$ and
$\beta^*_2=\beta^*_3=\beta^*_4=\vec{0}$, and observed responses $y_1,\ldots,y_n$ are a realization of $Y_1,\ldots, Y_n$ 
defined by \eqref{MultiGen}.   We consider $(n,\delta) \in \{50,75\} \times \{1,3\}$. 

To evaluate the proposed two-step method, we investigate its ability to detect the correct response category grouping. 
It is possible to detect the correct number of groups but the incorrect structure: we call this \textit{incorrect}.  It is also possible 
that the method selects more groups than it should, and if two of
the groups were combined it would result in the correct structure: we call this \textit{one-step}.  
Turning parameters were selected from a subset of $\{10^{-10},10^{-9.98},\ldots 10^{9.98},10^{10}\}$ and we set $\Ls=\Ss
$.  Since this simulation study has only 4 response categories, it was
computationally feasible to compute the best AIC by searching over all possible response category combinations, which makes 
this exhaustive search a natural competitor.

In Table \ref{GFMRTab1}, we report the proportion of replications that the group structures of interest are selected for each $(n,
\delta)$
combination for the two-step method.  Table \ref{FullTab1} presents the results for the exhaustive search. These tables show 
that for every of $n$ and $\delta$, the two-step method correctly picks the true response categories groups
more than any other group structure.  In particular,  the proposed two-step method performs as well or better than exhaustive 
search for each $(n,\delta)$.  These results show improvement in model selection performance when compared to
the simulation results from Section \ref{predpre}, suggesting
that if model selection is the interest the two step approach should be used.  

\begin{table}
\caption{The fraction of the 100 replications specific group structures are selected for each $N$, $\delta$ combination using the two-step method . The label 
One-Step indicates that the correct group structure is still a possibility if the correct fusion was done with an additional 
combination.}
\label{GFMRTab1}
\begin{center}
\scalebox{.8}{
\begin{tabular}{c|cc|cc} 
  & $N=50$ &   & $N=75$ &   \\
  & $\delta=1$ & $\delta=3$ & $\delta=1$ & $\delta=3$ \\
\hline 
1 Group & 19/100 & 0/100 & 3/100 & 0/100 \\ 
\hline
2 Groups (Correct) & 58/100 & 71/100 & 80/100 & 83/100 \\ 
2 Groups (Incorrect) & 0/100 & 0/100 & 0/100 & 0/100 \\ 
\hline 
3 Groups (One-Step) & 20/100 & 21/100 & 16/100 & 15/100 \\  
3 Groups (Incorrect) & 0/100 & 0/100 & 0/100 & 0/100 \\ 
\hline 
4 Groups & 3/100 & 8/100 & 1/100 & 2/100 \\ 
\end{tabular} 
}
\end{center}
\end{table}

\begin{table}
\caption{The fraction of the 100 replications specific group structures are selected for each $N$, $\delta$ combination when all 
possible response category combinations were used as candidate models.}
\label{FullTab1}
\begin{center}
\scalebox{.8}{
\begin{tabular}{c|cc|cc} 
  & $N=50$ &   & $N=75$ &   \\
  & $\delta=1$ & $\delta=3$ & $\delta=1$ & $\delta=3$ \\
\hline 
1 Group & 0/100 & 0/100 & 0/100 & 0/100 \\ 
\hline
2 Groups (Correct) & 46/100 & 59/100 & 62/100 & 82/100 \\ 
2 Groups (Incorrect) & 25/100 & 4/100 & 14/100 & 0/100 \\ 
\hline 
3 Groups (One-Step) & 28/100 & 29/100 & 24/100 & 17/100 \\  
3 Groups (Incorrect) & 0/100 & 0/100 & 0/100 & 0/100 \\ 
\hline 
4 Groups & 1/100 & 8/100 & 0/100 & 1/100 \\ 
\end{tabular} 
}
\end{center}
\end{table}

\section{Election Data Example}

We analyze the dataset \texttt{nes96} found in the CRAN package
\texttt{faraway} \citep{farawaypack}. The response variable,
self-identified political 
affiliation of voters, has 7 levels: strong Democrat, weak Democrat, independent Democrat, independent, independent 
Republican, weak Republican, strong Republican. The explanatory variables are voter education level (categorical with 7 levels),
voter income (categorical with 24 levels), and voter age (numerical).   An investigation 
into both model selection based on validation likelihood tuning parameter selection and model selection based on using the two-step method was performed using  $\Ls=\Ss$ and $\Ls=\{(1,2),\ldots, (6,7)\}$.  

Model selection for GFMR using validation likelihood tuning parameter selection was performed using 5-fold cross validation 
selecting the tuning parameter from the set \\
$\{10^{-10},10^{-9.98},\ldots, 10^{9.98},10^{10}\}$. In Table \ref{ElectionData1}, we present 
the response category combinations recommended by the GFMR regression coefficient estimates when $\Ls=\Ss$ and $\Ls=\{(1,2),\ldots, (6,7)\}$. The results show for the case when $\Ls=\Ss$, GFMR does not combine any response categories. When $\Ls=
\{(1,2),\ldots, (6,7)\}$, the results show that independent Democrats, independent Republicans, and independents have the 
same estimated regression coefficient vectors.

We also show the results from the two-step approach proposed in Section \ref{TuneReg}.  Both choices for $\Ls$ resulted in a 
selected model with three response categories.  These response category groups are shown in Table \ref{ElectionData}.

 \begin{table}
 \begin{center}
 \caption{The response categories found by  GFMR with tuning parameter selection using validation likelihood on the 1996 United States election data.}
 \label{ElectionData1}
 \scalebox{1}{
\begin{tabular}{c|c|c}
 Group & $\Ls=\Ss$ & $\Ls=\{(1,2),\ldots, (6,7)\}$\\ 
 \hline 
 1 & Strong Republican & Strong Republican\\
 \hline
 2 & Weak Republican & Weak Republican\\
 \hline
 3 & Independent Republican & \begin{tabular}{c}Independent Republican \\ Independent\\ Independent Democrat \end{tabular}\\
 \hline
 4& Independent & Weak Democrat\\
 \hline
 5& Independent Democrat & Strong Democrat\\
 \hline
 6& Weak Democrat & \\
 \hline
 7& Strong Democrat& \\
 \end{tabular}
 }
 \end{center}
 \end{table}

 \begin{table}
 \begin{center}
 \caption{The response categories found by GFMR using the two step method for model selection on the 1996 United States 
 election data.}
 \label{ElectionData}
 \scalebox{1}{
\begin{tabular}{c|c|c}
 Group & $\Ls=\Ss$ & $\Ls=\{(1,2),\ldots, (6,7)\}$\\ 
 \hline 
 1 & \begin{tabular}{c}
 Strong Republican\\ Weak Republican\\ Independent Republican\\ Independent Democrat\end{tabular} & \begin{tabular}{c}Strong Republican\\ Weak Republican\end{tabular} \\ 
 \hline 
 2 & \begin{tabular}{c} Independent \end{tabular} & \begin{tabular}{c}Independent Republican\\ Independent\\ Independent Democrat\end{tabular} \\ 
 \hline 
 3 & \begin{tabular}{c} Strong Democrat\\ Weak Democrat \end{tabular} & \begin{tabular}{c}Strong Democrat\\ Weak Democrat 
\end{tabular} \\ 
 \end{tabular}  
 }
 \end{center}
 \end{table}
 
 \bibliographystyle{jcgs}
\bibliography{Paper}

 \end{document}